# Burstiness of Intrusion Detection Process: Empirical Evidence and a Modeling Approach

Richard Harang, Alexander Kott

*Abstract*—We analyze sets of intrusion detection records observed on the networks of several large, nonresidential organizations protected by a form of intrusion detection and prevention service. Our analyses reveal that the process of intrusion detection in these networks exhibits a significant degree of burstiness as well as strong memory, with burstiness and memory properties that are comparable to those of natural processes driven by threshold effects, but different from bursty human activities. We explore time-series models of these observable network security incidents based on partially observed data using a hidden Markov model with restricted hidden states, which we fit using Markov Chain Monte Carlo techniques. We examine the output of the fitted model with respect to its statistical properties and demonstrate that the model adequately accounts for intrinsic "bursting" within observed network incidents as a result of alternation between two or more stochastic processes. While our analysis does not lead directly to new detection capabilities, the practical implications of gaining better understanding of the observed burstiness are significant, and include opportunities for quantifying a network's risks and defensive efforts.

## I. Introduction and Motivation

### A. Motivation

We explore approaches to modeling the detection process of cyber infections, i.e., presence of malicious software – malware – on a computer network. Often, this process is performed by a specialized organization – Managed Security Service Provider (MSSP) [1] – that monitors computer networks, analyzes the information obtained from the network, detects intrusions and activities of malware on the network, and reports such detections to the operators of the network who then take measures necessary to recover from the intrusion. As this research is based on empirical data provided to authors by a MSSP covering several large, non-residential networks, hereafter we use the term MSSP when referring to a network's cyber defenders.

For the purposes of this paper, we describe the process

This paper was submitted for review 3 November 2016. This work was supported in part by the U.S. Army Research Laboratory.

Richard Harang is with the U.S. Army Research Laboratory, Adelphi MD (e-mail: richard.e.harang.civ@mail.mil).

Alexander Kott is with the U.S. Army Research Laboratory, Adelphi MD (e-mail: alexander.kott1.civ@mail.mil).

executed by a MSSP organization in a limited and simplified fashion as follows: (1) the MSSP deploys one or more network sensors, i.e., hardware and software that collects information from the network traffic, on the computer network of its customer; (2) the sensor monitors the traffic to and from the customer's network; (3) the sensor sends alerts and information captured from the network (possibly including full-packet capture of selected traffic) to the MSSP analysis facility; (4) the MSSP's human analysts and automated analytical tools process the information arriving from the sensor; (5) when a MSSP analyst, having analyzed the available information, concludes that malware is operating on the customer network, the analyst prepares an incident report and sends it to the customer's network administrator as well as to pertinent authorities. Thus, the process of monitoring and analysis yields a series of incident reports each of which documents a detection of an infection. MSSP analysts strive to submit each report as soon as possible after detecting an infection.

The major findings of this paper are that the timing of when the reports are filed is not random and uniformly distributed over an interval (i.e., a Poisson process) but instead exhibits significant burstiness, and furthermore that this burstiness can be modeled well by a simple two-state model. While this lends no new insight as to *how* to find novel intrusions, understanding and modeling the dynamics of burstiness in intrusion detection would help anticipate the variability of detections over time using established intrusion detection methods. One obvious benefit that would accrue from such a model, if calibrated for a given network, would be to enable a manager of MSSP operations to project the workload and allocate and schedule the efforts of cyber analysts and other resources in a more effective manner (compare with similar arguments in [2]).

### B. Malware Lifecycle

To introduce key elements of the phenomena explored in this paper, we describe here the lifecycle of malware. For the purposes of this paper, the first event of interest in the lifecycle of malware is when a cyber attacker successfully deploys a malware on the network defended by a MSSP. Following the deployment, the malware and its controller begin executing activities, such as beaconing or downloading additional malware.

If the MSSP possesses means such as a signature or a behavioral rule to detect the activities of the malware, the detection and removal of the malware typically occurs rapidly, and thus has a very limited opportunity to cause harm; in this paper we refer to such relatively benign events as "soft" intrusions. Alternatively, if the MSSP does not possess means to detect activities of a particular type of malware – often a novel malware or a variation on existing malware that evades existing signatures – the cyber attacker is able to execute further stages of exploitation and expansion, such as delivering additional malware of the same or different type into the network, propagating further through the network, encrypting data as part of a ransomware attack, or exfiltrating information that interests the cyber attacker; in this paper we will refer to such intrusions as "hard" intrusions. As the malware performs these activities, some may be noticed by cyber defenders as suspicious. Eventually, such suspicious indicators would accumulate to a point where detailed analysis will reveal activities of a malware, and a corresponding means of detecting the malware, e.g., a signature or a detailed behavioral pattern. Here we use the term signature to refer to any means that rapidly reveal activities of a given type of malware.

Having obtained the new signature, MSSP analysts use recent as well as previously stored records of network traffic and of host-based activities to seek information that matches the signature, often resulting in rapid detection of multiple related malware instances on the network, and rapid generation of multiple intrusion reports. Because such intrusions tend to reflect more prolonged and therefore potentially more harmful activities, it is worthwhile to differentiate these hard-to-detect incidents ("hard" incidents) from the relatively easy-to-detect and therefore less harmful "soft" incidents.

Therefore, to oversimplify, the process of detecting intrusions can be approximated for modeling purposes by a process driven by two broad types of malware and the corresponding two types of detection processes: the process of soft intrusions that yields a continuous stream of easily detected and short-lived individual infections, and a more involved process of periodic detection of potentially long-lived, extensively propagated and dangerous hard infections that result in multiple intrusions. In practice, however, it can be difficult to distinguish between these two types, and there exists a spectrum of intermediate variations between these two.

Even an effective MSSP organization that has a significant experience detecting malware on a given network recognizes that much of the malware lifecycle remains hidden. Even after uncovering and reporting an infection, a MSSP analyst is rarely certain about how long the reported malware has been present on the network and what have been its past activities. Although a detailed cyber-forensic investigation may shed light on such questions, it is conducted relatively infrequently because of the associated expense and need for scarce forensic expertise.

*C. Temporal Features*

As we demonstrate in this paper – to our knowledge for the first time with respect to multiple organizations and several statistical techniques – the process of detecting malware, like many other processes, occurs in a "bursty" fashion. Burstiness refers to a tendency of certain events to occur in groups of relatively high frequency, i.e., with short inter-event time intervals, followed by periods of relatively infrequent events. Many processes exhibit some form of burstiness, including anthropogenic processes such as sending emails [3] market trading [4], watching movies [5], listening to music [6], and playing games [7], traffic in computer and communications networks [8], [9], and many natural events such as earthquakes [10] and neuronal firing [11].

Unlike sequences of events formed by strictly memoryless random processes – which may also happen to display apparent clusters of more frequently occurring events purely on the basis of chance – a true bursty process cannot be adequately described by a Poisson process with a corresponding exponential distribution of statistically independent inter-event intervals. Thus, bursty processes may be characterized [37] with respect to the degree to which they deviate from a Poisson process's inter-event time distribution. Furthermore, the nature of mechanisms that produce burstiness in different processes differs widely and is often uncertain and subject to disagreements among researchers, see e.g., [13].

In the later sections of this paper, we use several statistical tests, especially the Kolmogorov test [14] and the K-statistic of Ripley [15] to confirm burstiness from several perspectives and across several different sets of infections. We then investigate memory as well as burstiness properties of these processes, as described in [37], through the lens of a simple two-state mechanism that can be shown to produce similar properties in contexts other than cyber security [16]. We conclude that the nature of the process under consideration is more reminiscent of natural processes such as earthquakes than of anthropogenic processes such as sending emails. Unlike anthropogenic processes, the process of intrusion detection events exhibits strong memory, and the likely mechanism of the burstiness in intrusion detection is reminiscent of the integrate-and-fire [17] [18] or similar threshold [16] phenomenon: knowledge about a new malware accumulates to the point until it becomes actionable and enables analysts to recognize a particular type of intrusion that was previously difficult or impossible to find. At that point, the analysts are able to rapidly recognize a number of pre-existing intrusions within a network(s) under their care and produce multiple reports in rapid succession.

*D. Data and Sources*

In this research, we use records of reports produced by a MSSP organization that serves multiple organizational customers. Each report in our data corresponds to a confirmed finding of active malware on a monitored computer, detected by automated tools and verified by a human analyst; for this analysis, we have ignored other events commonly of interest to an MSSP -- such as misconfiguration of devices -- in order to focus on malware-related events.

The data used in this research consist of five sets for the purposes of identifying burstiness, and three of the five sets were used for the purposes of constructing a model. Although a number of other datasets are available from the same MSSP, these five were selected because they have the largest number of intrusion reports over the period of observation for which consistent records are available. Each set refers to one network controlled by one organization –a customer of the MSSP – for a total of five organizations. The organizations differ significantly in their business types, geographic locations, culture, personnel, operations, cyber-related policies, and likely threats. Each set was observed over multiple months, and the rates of intrusions differ widely between the networks. Table I shows number of records for each network we studied. Table II, column "Inter-event time" shows that rates of intrusions differ significantly between networks, depending on the nature of the business and the defenses of the organization that uses each network.

A set consists of multiple records. Each record represents one report of an intrusion and contains, at a minimum, the following information which we use in this research: {Identifier of the Network (in our case: A, B, C, D, E, see Table I); Date of Report Submission}. The report itself can be rather voluminous and describes details of the intrusion, the evidence that supports the conclusions of the analyst, and recommended actions to resolve this intrusion and to mitigate further damage. However, the content of the report is not the subject of this research; we focus here only on the timing of a report submission, for each of the networks. In all cases, reports refer to situations where an actual malware operated on the network, as opposed to other types of reportable cyber incidents.

Here, the difference between alerts and reports should be mentioned, because these are occasionally confused. An alert is generated by an automated intrusion detection tool when it detects a suspicious event; the volume of alerts is very high and the overwhelming majority of alerts are false positives. An incident report is generated by a human, qualified MSSP analyst after a due process of analyzing and correlating a large number of alerts as well as investigating other pertinent information; reports are relatively few and are generally believed to be true positives.

## II. PRIOR WORK

Two distinct bodies of prior work are related to this paper. First, we examine prior work on detection of bursting behavior in various contexts. The most prominent of these is detection of spatially clustered events, and we review methods for adapting these techniques to our problem. Next, we review work that examines various properties of bursting behavior and memory under the assumption that they are known to exist in the model. Our analysis follows a similar trajectory, first examining the time series of intrusion detection reports to demonstrate the presence of bursting behavior, and then a more detailed analysis of it using tools specialized to the task of examining burstiness. It is worth emphasizing that we focus on an empirical analysis of the intrusion detection process using an existing intrusion detection toolset. It is not the purpose of this work to propose a new intrusion detection technique.

### A. Approaches to analysis of clustering and burstiness

Analysis of bursting activity may be viewed as a special case of clustering analysis, restricted to a single temporal dimension rather than two spatial dimensions. While surprisingly few methods directly address the problem of detecting and testing for bursting, a wide variety of methods have been developed for detecting and analyzing spatial clustering, particularly in the geostatistical literature (see, e.g., [19], [20], [21], [22], [23]), where the detection problem often involves rare events such as uncommon diseases hypothesized to be precipitated by some common local cause [24], [25], [26], [27]. Due to the complications involved in precisely localizing the occurrence of a specific medical incident, much of the recent work in that area focuses on mitigating the bias inflicted by the need to aggregate over complex, irregular, and arbitrary spatial domains (e.g., geographic counties) [19], [22], [23], [27], introducing complications that are not germane to our problem, in which we have exact observations of arrival times of events.

Cluster analysis can be broadly split into two main categories [24]: "focused" and "general." In "focused" cluster detection, a potential cluster is tested for statistical significance. In "general" cluster detection, the isolation of specific clusters is set aside in favor of simply determining if the overall distribution of points appears to be roughly uniform or exhibits clustering. We focus in this paper exclusively on the "general" cluster detection problem: determining whether or not the process of network security incidents exhibits overall bursting behavior – clustering with respect to time – rather than determining whether particular sequences of events correspond to clusters.

Our problem is made simpler than many of those presented above due to the single dimension of interest (time), and the various well-known properties of the null hypotheses we investigate (homogenous Poisson process in a single dimension). In addition, the availability of exact arrival times allows us to avoid using the "general" cluster detection algorithms that operate on cluster centroids rather than exact location, and instead use methods that are designed to take advantage of the extra precision afforded by the exact arrival times (as opposed to discretized centroids). The methods we focus on are the Kolmogorov test [14] to examine the simplest null hypotheses with respect to arrival times, and the K-function of Ripley [15] with a modification that is similar to Besag's rescaling in the form of the L-function [25].

### B. Prior work on burstiness

An extensive literature explores burstiness in a wide range of processes other than intrusion detection process; [16] and [37] are two examples of that literature. With respect to intrusion detection, a number of authors explore bursts from a very different perspective than in this paper. Many researchers [28], [29], [30], [31], report intrusion detection techniques that use bursts of network traffic, bursts of connection requests, or bursts of login attempts as indicators of malicious activities. Similarly, bursts of messages may be indicative of spam campaigns [32]. Others investigate the negative impact that bursts of network traffic have on the accuracy of anomaly detection [33] or on the possibility that the intrusion detection tools may be unable to cope with an extremely high rate of alerts when they arrive in a major burst [2], [34]. Others propose effective approaches to handling and aggregating bursts of alerts [33], [35]. These works, however, do not address the burstiness in the intrusion detection process itself.

Very little prior work has been directed specifically to the burstiness of the intrusion detection process. A single paper [36] notes that infections exhibit burstiness and use a technique called Allan deviation to demonstrate this fact in application to one large set of infections observed on a network. Unlike prior work, in this paper we demonstrate, with strong empirical evidence that involves data from multiple organizations and use of several statistical techniques, the burstiness in the process of detecting infections effected by cyber threats against networks of large organizations. We also propose – for the first time, to our knowledge – a hypothetical mechanism for explaining the burstiness and a model that produces plausible parameters when fitted to the observed values.

## III. EVIDENCE OF BURSTINESS

To evaluate the presence of bursting behavior in the data obtained from the MSSP customer organizations (as described above), we extracted the time series of intrusion times ("events") and subjected them to the following tests. Initially, we perform Kolmogorov testing on both the arrival times and the inter-event times (lags between successive events) under a simple null hypothesis. We then use the tests of [15] and [37] to evaluate, in turn, the temporal distribution of the discrete events and the distribution – both with and without ordering – of the inter-event times, to evaluate the two aspects of the data separately. We provide parametric p-values, where possible and Monte Carlo results otherwise. Event counts for the data over the collection interval are provided in Table I, below. Details on the exact timing of events and the interval over which they were collected are beyond the scope of this paper.

As the number of observations varies between networks (see Table I), we emphasize the use of formal statistical methods with significance testing, which controls for the possibility of spurious results due to a combination of chance and a lack of data.

TABLE I
DATA SUMMARY

| Network | Number of events |
|---------|------------------|
| A | 51 |
| B | 916 |
| C | 77 |
| D | 168 |
| E | 718 |

### A. Kolmogorov tests for arrival and inter-event times

Here we present the first test of burstiness found in our data. The simplest null hypothesis is simply that of a homogenous Poisson process over the observed time interval. Formally, we have some constant $\lambda$ such that for any continuous interval $T = (t_0, t_1)$, the number of events $X_T$ in that interval is distributed as a Poisson random variable with parameter $\lambda \times (t_1 - t_0)$. The various distributional properties of this stochastic process are well known (see e.g. [38]), and may be directly tested for statistical significance via the Kolmogorov test.

Specifically, for an observation from a homogenous Poisson process taken on a bounded interval, the following three properties must all hold:

1. Conditional on the number of events within the interval, the arrival times of those events are uniformly distributed on the interval.
2. The inter-event times between successive events are independently and exponentially distributed.
3. For any given partitioning of the interval, each sub-interval $(t_i, t_{i+1})$ has a number of events that are also Poisson distributed, with parameter $\lambda \times (t_i, t_{i+1})$

We examine the first two of these properties and conclude that they do not hold, and as such, the null hypothesis of a homogenous Poisson process cannot hold (examination of the third is redundant in the face of the first two not holding, and leaving it aside allows us to avoid the more complex problem of examining all possible partitions of the data).

TABLE II
RESULTS FROM KOLMOGOROV TESTS

| Network | Count | Arrival time (test stat, p-value) | Inter-event time (test stat, p-value) |
|---------|-------|-----------------------------------|---------------------------------------|
| A | 51 | 0.2062, 0.02224 | 0.1455, 0.2180 |
| **B** | **916** | **0.08115, 0.00001074** | **0.2037, <1.0e-12** |
| C | 77 | 0.1541, 0.04608 | 0.1680, 0.02415 |
| **D** | **168** | **0.1679, 0.0001304** | **0.1802, 0.00003238** |
| **E** | **718** | **0.08177, 0.0001266** | **0.2166, <1.0e-12** |

Table II shows the results of testing against the null hypothesis of a homogenous Poisson process for the five networks evaluated. Results significant at a level of $p \leq 0.001$ have been emphasized in bold font. For the parameterization of null distribution for the inter-event time, the maximum likelihood estimator (the mean number of events per unit time) was used. While the results do show that for two of the networks (A and C) we clearly cannot reject the null hypothesis of a homogenous Poisson process, there are important caveats.

First, we reiterate that a failure to reject the null hypothesis does not immediately entail that the null hypothesis is true; simply that it was not falsified. We note that there appears to be a loose correlation between the count for a network and the p-value observed; this suggests that we may simply have accumulated insufficient data to falsify the hypothesis of homogeneity, particularly for network A.

Second, and more importantly, even a statistically significant result in these tests does not directly indicate clustering, as it does not exclude other alternatives such as an inhomogeneous Poisson process with a piecewise constant intensity function. It simply demonstrates that the hypothesis of a homogenous Poisson process is – with high likelihood – incompatible with the observed data.

Finally, although evidence of burstiness is strong for at least some of the networks we have investigated, it does not mean that burstiness is always present, or can be observed in all networks and in all intrusion detection processes. In other words, we show that burstiness is present in some cases, but not necessarily in all cases. It is entirely possible that there may exist classes of networks, and/or associated intrusion processes, and intrusion detection processes where burstiness is either absent in principle, or cannot be detected from the available data. As noted above, all the networks in this study bear stronger resemblance to corporate or institutional networks than to e.g. residential or academic networks; as such the conclusions that we reach may not apply to such networks. The fact that we observe such strong evidence of bursting in 3 of the 5 networks we have examined, however, leads us to suspect that this phenomenon is prevalent in many if not most high-volume networks. Identifying conditions under which burstiness exists, and can be observed, is a topic of future research.

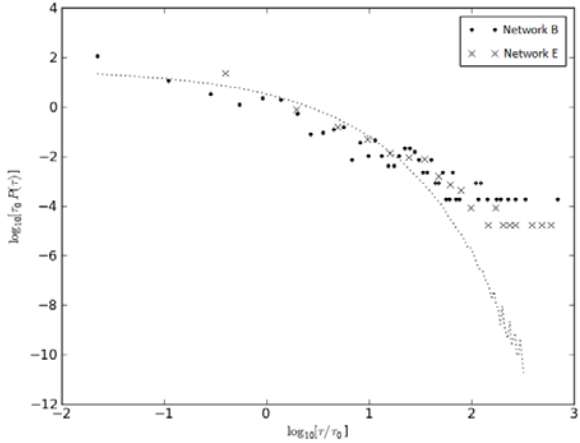

Fig. 1. Plot of inter-event distributions for networks B and E (the two largest data sets with the most obvious bursting behavior), normalized by average rate $\tau_0$ to lie on same scale; the distribution for an exponential distribution is given as a dashed line. See Figure 2A in [37] for comparison. The points lying above the dashed line indicate a "fat tail" of longer inter-event times than expected under an exponential distribution of inter-event times that corresponds to a Poisson process.

### B. Ripley's K-function

We now perform the second test of busrtiness; we explore the question of clustering in the temporal domain more directly by examining the distribution of arrival times using a variant of Ripley's K-function [15] adapted to a single dimension. Ripley's K-function is defined as

$$K(t) = \frac{1}{\lambda} E\big[\#\{X_j: |X_i - X_j| \le t\}\big]$$

where $X_i$ and $X_j$ are the arrival times of the $i^{th}$ and $j^{th}$ event, respectively. The $\#\{A\}$ denotes the cardinality of set $\{A\}$, and $E$ is the expected value operator, thus we compute the expectation (with respect to $X_i$) of the number of events $X_j$ that occur within some interval $t$ of $X_i$. This value is then normalized by $\lambda$, the average intensity (events per unit measure) of the total process.

Intuitively, the K-function estimates the tendency of the process that generates events $X_1, X_2, ...$ to cluster within some smoothing window $t$. If the data exhibit clustering within some radius $t$, then the value of the K-function in the vicinity of $t$ will be elevated, as the expected number of events within a span of $t$ from any arbitrary event $X_i$ will be above $\lambda$. The values of $t$ for which the K-function remains elevated provide an indication of the scale at which the bursting or clustering occurs. The K-function is most commonly used in geostatistics, where the measure of the Poisson process is defined on a two-dimensional domain. As our domain is different, we derive comparable statistics for our domain.

First, define the number of events within some window of size $t$ of some given time point $x_i$ by $C(x_i, t)$, which we may calculate as $C(x_i, t) = \#\{X_j: |x_i - X_j| \le t\}$ where $X_j$ ranges over the set of all other events. Note that, under the assumption of a homogenous Poisson process, we have that $C(x_i, t) \sim \text{Poisson}(2\lambda t)$. It immediately follows that $K_{mod}(t) = \frac{C(x_i,t) - 2\lambda t}{\sqrt{t}}$ will have zero mean and variance $2\lambda$. As the mean and variance are finite, we may apply the central limit theorem to find that $\frac{1}{N}\sum \frac{C(x_i,t) - 2\lambda t}{\sqrt{t}}$ will be asymptotically normally distributed with mean $0$ and variance $2\lambda/N$ for sufficiently large number of observations $N$.

While precise details are beyond the scope of this paper, we note the following results:

1. We observe significant indications of clustering under the K-statistic (p-value of <0.001, using the Gaussian approximation above) for all five time series when considering intervals on the order of 1-4 days.
2. For networks B, D, and E, where we have a larger number of observations, but not networks A and B, we observe clustering over even larger windows, again suggesting a correlation between the amount of available data and our ability to detect clustering.

### C. Inter-event analysis using burstiness and memory parameters

Our third test of burstiness involves an analysis of the distribution of inter-event times. We apply the method of [37] to the data to calculate clustering coefficients for all five networks. In particular, we parameterize bursting behavior in terms of a 'memory' parameter $\mu$ – loosely, the degree to which there is serial autocorrelation in the event interarrival times – and 'burstiness' $\Delta$ parameter, which characterizes the degree and manner in which the statistics of interarrival times differs from that suggested by a Poisson process. Due to the discrete nature of our data, generating a continuous density estimate for the inter-event times was impractical; we instead binned the data in $\sqrt{n}$ bins, calculated the corresponding interval probability directly under an assumed exponential distribution via the cumulative distribution function (CDF), and took the differences across intervals. As statistical tests are not developed in [37] due to the difficulty of clearly identifying the distribution of their test statistic, and are likely not feasible due to our transformation of the data, we instead apply a simple Monte Carlo test (not to be confused with the MCMC methods we use to fit the model later in the paper) in which the null hypothesis of a homogenous Poisson process with the MLE rate (as above, the mean number of events per unit time over the length of the data) was used. Ten thousand sample trajectories were generated for each network, and memory $\mu$ and burstiness $\Delta$ were calculated for each trajectory separately; upper and lower 95% limits were obtained from the empirical distribution thus generated and are reported in Table III. An asterisk denotes that the observed value fell outside of the 95% range of values observed in the Monte Carlo trials, suggesting that the observed value displays statistically significant burstiness and memory compared to a homogenous Poisson process generated from the same rate parameter.

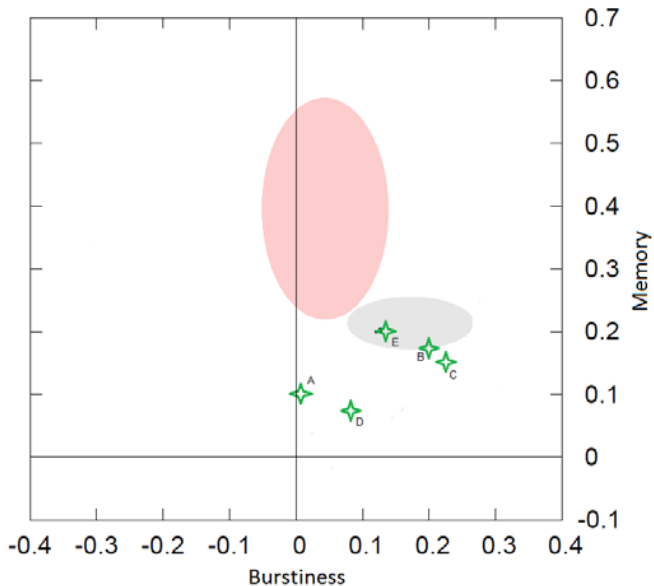

Fig. 2. Reproduced from 3b Goh and Barabasi [37] with our data superimposed as 4-pointed stars and labeled by network. According to [37] the region highlighted in red is associated with "human activity" (email, library loans, printing); the area in gray is associated with natural phenomena (rainfall, earthquakes). Our data trend towards the gray area, i.e., natural phenomena.

The two largest data sets (B and E) with the most obvious bursting behavior show significance under the Monte Carlo test (as the observed value for the data is outside the range of values observed in the most common 95% of the simulated trials). Figure 1 displays the plot of the estimated density for the inter-event times of network B and E against an exponential distribution, normalized to a common scale (see also Figure 2A in [37]). The long-tailed nature of the inter-event distribution (relative to an exponential distribution corresponding to the homogeneous Poisson process) for both networks is readily visible. While the log-scaling reduces the relative counts in the smaller bins, there is some indication for both networks that a slight excess of shorter waiting times is observable as well.

It is interesting to note that set D (which had a highly significant p-value under the Kolmogorov test) does not exhibit significance under this metric, while C (which had only a marginally significant p-value under the Kolmogorov test) does show significance under the Monte Carlo test for the burstiness test statistic. This suggests that network E may not exhibit strong bursting behavior, despite both arrival and inter-event times not following the distributions associated with a Poisson process, while the marginal statistical significance for network C under the Kolmogorov test may be simply a function of sample size.

TABLE III
BURSTINESS AND MEMORY PARAMETERS

| Network | Observations | Burstiness ($\Delta$) | Memory ($\mu$) | 95% MC Range for $\Delta$ | 95% MC Range for $\mu$ |
|---|---|---|---|---|---|
| A | 51 | 0.1015 | 0.03103 | -0.1865 to 0.1758 | -0.2592 to 0.2599 |
| B | 916 | 0.1726* | 0.2022* | -0.06856 to 0.06633 | -0.06307 to 0.06557 |
| C | 77 | 0.1551* | 0.2281* | -0.1600 to 0.1428 | -0.2092 to 0.2170 |
| D | 168 | 0.0866 | 0.0890 | -0.1215 to 0.1130 | -0.1479 to 0.1513 |
| E | 718 | 0.2002* | 0.1330* | -0.07417 to 0.07176 | -0.07160 to 0.07299 |

In Figure 2, we reproduce Figure 4b of [37] with our data superimposed in the form of green four-pointed stars. It is worth remarking that all three networks that (B,C, and E) displayed significant burstiness and memory values under the Monte Carlo test appear to be very nearly in the same range as the "natural phenomena" (earthquake and precipitation records) observed in [37]. While the remaining two points (A and D) appear to cluster separately, numerical experimentation (data not shown) suggests that the magnitude of both parameters can be extremely variable at low sample sizes; this may be observed in Table III by comparing the number of observations with the 95% windows for $\Delta$ and $\mu$. Additional data are required to determine if these two points truly are forming a separate cluster indicative of a separate mode of network infection that does not show characteristics of a natural process, or if they will – given sufficient data – move to an existing cluster. However, the grouping of the process of some network security incidents with natural events rather than either anthropogenic events or purely random events nevertheless suggests the intriguing possibility that some fundamental law may serve a limit to the process of exploiting networks under some circumstances.

TABLE IV
SUMMARY OF TESTING RESULTS

| Network | Observations | Kolmogorov test | K-statistic | Burstiness MC test | Memory MC test |
|---|---|---|---|---|---|
| A | 51 | - | + | - | - |
| B | 916 | + | + | + | + |
| C | 77 | Marginal + | + | + | + |
| D | 168 | + | + | - | - |
| E | 718 | + | + | + | + |

Combined with the results obtained from the Kolmogorov testing and examination of the K-statistic, above, we conclude that there is extremely strong evidence of clustering in networks B, and E, somewhat less strong but still very compelling evidence for burstiness in network C, and marginal evidence for networks D and A. While all networks displayed some degree of temporal clustering, and only network A had a clearly non-significant Kolmogorov test result, the burstiness and memory parameters are less clearly distinguishable for networks D and A (see Table IV that summarizes all results of tests, with "+" for evidence for burstiness and "-" otherwise.). It should be noted that the highlighted regions are not indicators of e.g. statistical significance, but rather approximate regions for particular types of processes identified in Goh and Barabasi [37]. The main item of interest is that – to the degree that the points deviate from memoryless/non-bursting processes – they do so in a manner that is more reflective of natural processes than human-generated ones.

We once again note that there seems to be some degree of correlation between the number of observations available for a network and the power of the associated tests. One may conjecture that burstiness can be detected only when the number the number of observations is sufficiently high. Identifying conditions (such as the required number of observations) under which burstiness can be detected is a topic of future research.

*D. A Hypothetical Mechanism of the Burstiness*

We hypothesize that the mechanism of burstiness has to do with what we call a threshold of analyst knowledge. As conjectured in [16], the common element of various bursty processes – even if very different in nature – is a threshold mechanism, i.e., events occur infrequently until some domain-specific quantity accumulates to a threshold value, at which point the events "burst out" at a high frequency.

Intriguingly, cyber analysts recalled episodes when multiple discoveries of intrusions (and corresponding reports) are made after arrival of a crucial piece of new information about a previously unknown malware behavior or characteristic. This new information enables analysts to recognize a particular type of intrusion that until then was difficult or impossible to find. At that point, the analysts are able to rapidly recognize a number of pre-existing infections within a network(s) under their care and produce multiple reports in rapid succession.

Clearly, this is a kind of threshold mechanism, where the available knowledge of analysts must reach a certain critical value before enabling a burst of new discovery events. As shown in [37], processes with obvious threshold mechanisms, such as earthquakes, exhibit strong memory (see Figure 2 for illustration). In our data, we notice that the intrusion detection process also tends towards higher values of the memory parameter (see points E, B, and C in Figure2). This could be interpreted as a support to the hypothesis that the busrtiness in the intrusion detection process is also associated with a threshold mechanism. Although we do not know if this mechanism is the only, or even the primary one responsible for burstiness, it can serve as a working hypothesis and a motivation for the modeling approach we consider next.

IV. MODELING THE BURSTY PROCESS

Having established that burstiness is clearly present in networks with a larger number of observations, and at least plausibly present in the rest, we turn our attention to attempting to model this phenomenon.

We examine a general class of models that is a slight modification of that presented in [16]. The discrete model presented in [16] considers the system to be in one of two states – "normal" or "excited" – at all times. In the context of our domain, the normal state is when the MSSP detects only soft intrusions, while the excited state is when the MSSP analysts obtain a signature for a new type of malware and begin discovery of hard intrusions – often multiple – in rapid succession, leading to a burst of reports. In both states, the waiting time between successive events is generated by a long-tailed discrete distribution, in which the probability of waiting for one additional unit of time conditional on having waited for time $t_{ie}$ since the previous event is given by:

$$f(t_{ie}) = \left(\frac{t_{ie}}{t_{ie}+1}\right)^{\mu_s},$$

where $\mu_s$ is the reinforcement parameter for the current state.

When the system is in the normal state, it waits for the time induced by $f$ prior to emitting another event, and then with some fixed probability transitions to the excited state. The excited state introduces another long-tailed function $p(n)$ – with form essentially identical to $f(t_{ie})$ above – which governs the probability of emitting another event in the excited state conditional on having emitted $n$ events so far. Note that the value of $f$ approaches 1 as the argument grows, indicating a self-reinforcing property by which the likelihood of waiting additional time before emitting another event grows as the time since the last event increases. This behavior is modulated by the parameter $\mu_s$, which for values greater than 1 will reduce the probability of waiting an additional unit of time across all inter-event times.

While this two-state model is a very coarse representation of the underlying process of intrusion detection and does not, for example, distinguish between different varieties of malware or categories of detection tools, models of this kind have been extensively studied in the context of bursty systems [16], and are thus reasonable to use as a model of the phenomenon we are attempting to study.

The physical intuition of the two-state model is as follows:

in our interpretation, the normal state is the state in which the network defenders observe the easy-to-detect intrusions, using the available detection techniques (see the discussion of the hypothetical mechanism of burstiness in section III.D). Meanwhile, a malware for which the detection techniques is yet unknown keeps accumulating within the network. Eventually, sufficient number of observations and other knowledge about the hard-to-detect type of malware leads to the discovery of a detection technique for the malware. This creates the excited state, i.e., the state in which the defenders of the network are actively uncovering multiple "hard" infections using the newly found technique. This corresponds to the "feast or famine" pattern of activity anecdotally reported by MSSP analysts, in which periods of normal activity are interrupted by bursts of high activity when novel infections are discovered. A physical analogy could be drawn to earthquakes -- explored using the two-state model in [16] -- in which strain around the fault surface (in our example, accumulation of infections and associated indicators) accumulates to some critical point (in our case, the point of discovering a detection technique), whereupon it is discharged through a series of earthquakes and aftershocks (in our case, detections of malware).

In order to model the continuous-time nature of our data, we further modify the discrete model of [16] to a continuous time version. Due to the nonconvergence of the integrals of functions of the kind investigated in [16] when considered on the non-negative real numbers, we moderate the long-tailed behavior by making the incremental probabilities constant – i.e., exponential waiting times between events. The state remains continuous; analogously to our constant-increment modification for waiting times, we make the incremental state probabilities constant as well, thus inducing a geometric distribution on the number of events fired from the excited state.

The simplified system with continuous time and discrete state thus obeys the following dynamics: the inter-event times are exponentially distributed with parameter $\lambda_s$, while the number of emissions in each state are geometrically distributed with parameter $p_s$, where $s$ indexes the current state. Given observable events, we can construct it as a hidden Markov model (HMM) in continuous time, where the state of the system (normal or excited) is hidden, and all other factors (most importantly inter-event times) are observed. In standard probabilistic graphical model notation [39], we write the following:

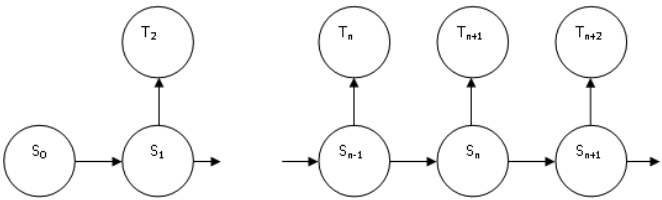

Fig. 3. Dependence diagram for the two-state model.

Where $S_n$ denotes the state after the $n^{th}$ emission event, i.e., in our case an issuance of an incident report, while $T_n$ denotes the time between emission event $n$ and $n-1$, and so the smallest $n$ for which $T_n$ is defined is $T_2$. We observe $\{T_n\}_{N=2}^N$, from which we attempt to infer the remaining parameters. Note the change in discretization between the model above and the model of [16]; while they consider a model that is discrete in both time and state, we allow the model to run in continuous time, and exploit the discretization of state and observed emission times to render the model tractable.

Such models are readily estimated by standard methods [40] however note that due to the symmetry between states, such systems may not have a unique parameterization. For this reason, when fitting the model to observed data (Section IV.) we use Bayesian methods with informative priors to ensure that the normal state is associated with lower rates of activity than the excited state.

V. FITTING THE MODEL

Having defined the two-state model we consider methods to fit the parameters of the model to the observed data. The simplicity of the two-state model, along with the strict dependence of the inter-event times on the (unobserved) state of the model, places it firmly within the class of hidden Markov models, for which a wide range of fitting approaches apply. We focus here on a Bayesian approach, and briefly derive a Gibbs sampling scheme for the model.

Under our continuous parameterization of the two-state model where $S_i=1$ denotes the excited state, with $\lambda_1$ the associated rate parameter (and similarly for $S_i=0$ for the normal state), we define $p_i$ as the probability of being in the excited state after the next event emission in state $I$ (i.e., the transition probability for the normal state, and the complement of the transition probability for the excited state), and $T_i$ the inter-event time between observation $i$ and $i-1$, estimating the posterior distributions for it through Markov Chain Monte Carlo (MCMC) [40] techniques is straightforward. The factorization of the joint likelihood becomes:

$$P(\{T_i, S_i\}_{i=1..N}, S_0, \lambda_0, \lambda_1 p_0, p_1)$$
$$= P(\{T_i, S_i\}_{i=1..N}, S_0 | \lambda_0, \lambda_1, p_0, p_1) P(\lambda_0) P(\lambda_1) P(p_0) P(p_1)$$

Where:
$$P(\{T_i, S_i\}_{i=1..N}, S_0 | \lambda_0, \lambda_1, p_0, p_1)$$
$$= P(S_0) \prod_{i=1}^N P(S_i | S_{i-1}) P(T_{i+1} | S_i)$$

(We suppress dependence on the parameters $\lambda_i$ and $p_i$ where it is clear from context.) Less formally, we can compute the total likelihood of any given sequence by first computing the probability of the initial state ($S_0$); the interarrival time of the first event ($T_1$) then depends upon that state (excited or normal), and the transition to the next state depends on $p_i$. By stepping through the model, we may compute the total likelihood of a given sequence of states and interarrival times given the state transition probabilities and the rate constants associated with each state, and thus find a parameterization (or set of parameterizations) that maximize the observed likelihood under this model. However, we only have observations on the $\{T_i\}$ values, and wish to find the posterior distribution:

$$P(\lambda_0, \lambda_1, p_0, p_1 | T_i) = \int_{\{S_i\}} \frac{P(\{T_i, S_i\}_{i=1..N}, S_0 | \lambda_0, \lambda_1, p_0, p_1) P(\lambda_0) P(\lambda_1) P(p_0) P(p_1)}{P(\{T_i, S_i\}_{i=1..N}, S_0)} \quad (1)$$

Or – again informally – we wish to estimate the distribution of the rate parameters and transition probabilities conditional on the observed data, which does not include information on the states $S_i$. Intuitively, we may guess that periods in which the interarrival time is generally large may correspond to a 'normal' state, and periods in which it is short may correspond to an 'excited' state; given those assumptions we may estimate both the rate at which the model emits events in both of those states, as well as the probability of transitioning from one state to the other after each event. By use of MCMC techniques we may formalize this intuitive approach into a strategy by which we may estimate the distribution of parameters we are interested in via sampling.

We have the assumed transition probabilities from excited to normal and back are given by:

$$P(S_{i+1} = 1 | S_i = j) = p_j$$
$$P(S_{i+1} = 0 | S_i = j) = 1 - p_j$$

And the interarrival times for a given state are generated by:

$$T_i | S_{i-1} \sim Exp(\lambda_{S_{i-1}})$$

For the unknown parameters, we use conjugate priors for simplicity:

$$p_i \sim Beta(1,1)$$
$$\lambda_0 \sim Gamma(1,2)$$
$$\lambda_1 \sim Gamma(3,2)$$

The prior on $p_i$ is uninformative, allowing us to avoid any strong bias on the posterior estimate. The priors on $\lambda_i$ are weak but informative to ensure that the excited component has a higher rate on average than the normal component. This avoids issues that can arise with bi-stable configurations when using Gibbs sampling.

This factorization makes it clear that conditional on the values of $T_{i+1}$, $S_{i-1}$, and $S_{i+1}$, we may calculate unnormalized likelihoods for $P(S_i = 1 | T_{i+1}, S_{i-1}, S_i)$ and $P(S_i=0 | T_{i+1}, S_{i-1}, S_i)$ with relative ease. As $S_i \in \{0,1\} \forall i$, these un-normalized likelihoods may be easily normalized and thus sampled from. Given the sequence of states $\{S_i\}$, the conditional posteriors for $p_i$ and $\lambda_i$ are even more straightforward to calculate and thus sample from. Finally, given the state assignments $\{S_i\}$ and the inter-event observation times $\{T_i\}$, we may re-estimate the conditional posteriors for $\lambda_i$ and sample from those as well.

This immediately suggests a Gibbs sampling strategy, in which we first sample each $S_i$ individually given the observation $T_{i+1}$ and the current samples values of $S_{i-1}$ $S_{i+1}$, $p_i$ and $\lambda_i$, and then re-sample the now conditionally independent parameters $\lambda_0$, $\lambda_1$, $p_0$, and $p_1$ conditional on the new sample for $\{S_i\}$. Note that the likelihoods for $S_0$ and $S_N$ require corrections for edge effects, however this correction is straightforward. By alternating these two sampling blocks, we may produce a MCMC estimate of the unknown parameters, including $\{S_i\}$. By ignoring the portions of the samples involving $\{S_i\}$, we effectively perform a stochastic version of the integration in equation 1, thus obtaining a sample from the posterior distribution of interest. For more complete details about Gibbs schemes in general, we refer interested readers to the excellent introduction provided by Gelman [40].

When fitting the data, we start a total of ten chains from random initial conditions, drew 5000 samples; convergence was assessed by the Gelman-Rubin diagnostic [41]. On the basis of this diagnostic, the first 1500 samples were discarded from each run as burn-in, and the remaining samples pooled into a single set of estimators for a total of 35000 samples. Detailed results and discussion are provided in section VI.

## VI. RESULTS

An example fit of data from a single network using the two-state model is shown in Figure 4. Loosely, in the top plot of Figure 4, each point denotes a combination of event rates in normal (x-axis) and excited (y-states) that might explain the observed data. Intuitively, the area of the plot with the highest density of the points represents the most likely "correct" combination of event rates in the two states. Similarly, in the bottom plot of Figure 4, each point represents the combination of switching probabilities that might explain the observed sequence of events: from normal to excited state (x-axis), and from excited to normal state (y-axis).

Due to the large number of samples, we have thinned the data by sampling every 35$^{th}$ point in order to obtain the 1000 samples shown in the figures; this was done solely to avoid clutter in plotting, and the full data set was used in numerical analysis. Rate parameters for the 'normal' and 'excited' states are jointly plotted in the top panel, along with histograms over marginal distributions, allowing us to examine both the joint and marginal distribution of these parameters.

We see that the rates for the normal state (x-axis) concentrate about a rate of approximately 0.007: up to an order of magnitude smaller than in the excited state (y-axis) which concentrated about a rate of 0.05. Recall that according to our hypothesis of section III-D applied to the two-state model (discussion in section IV), the interpretation of the normal state is that it represents a period when mainly easy-to-detect "soft" events are detected and recorded; and the interpretation of the excited state is that it represents a period when intrusion detection analysts obtain a new "signature" that enables detection of a number of pre-existing "hard" infections within a network(s) under their care and produce multiple reports in rapid succession. It should be noted that this fitting process does not tell us whether any particular intrusions are "hard" or "soft." Instead, it characterizes the overall process of intrusion detection observed for a given network, such as rates of detection reports and probabilities of switching for both states.

The switching probabilities ($p_0$ and $1-p_1$) are shown in the bottom panel. The trends are similar, with strong correlation between the switching probabilities in all three plots. The switching probability for the excited states (y-axis) have mode and mean values uniformly close to zero, indicating strong persistence in the excited states, while the switching probabilities in the normal state are much more diffuse, suggesting a wider range of uncertainty about the parameter.

Similar results can be observed across all organizations, and are consistent with visual inspection of the data: bursts – produced when the model transitions from a normal to excited state – tend to produce a larger number of events at a higher rate than in the normal state. Fewer events are produced in the normal state (reflected in the higher switching probability for that state), however the much lower rate parameter results in substantially larger time intervals between successive events in that state, with higher variance.

This two-state model is simple, and reflects an inherently limited bimodal process. Therefore, it may have difficulties in modeling, e.g., a new campaign of easily detectable malware which may raise the detection rate to an intermediate point. On the other hand, the model offers the ability to factorize it into a Gibbs sampler – as discussed in the previous section – and thereby makes it direct to fit and evaluate convergence of the fitting process. It also does not impose – in our experience with this model – excessive computational requirements to run in a generative fashion.

The fitting takes little time (e.g., on the order of minutes on a commodity desktop computer using 3GHz quad-core Intel processors), and can be accomplished even with a small number of data points (although the accuracy of the resulting parametrization may be doubtful). For all data sets, the two-state model yields plausible values of parameters with which it produces report-generation trajectories closely matching the actually observed data. It would be desirable to compare this model with alternative models, especially those proposed in prior work. However, to our knowledge, the prior work on burstiness in the intrusion detection process consists of a single paper [36]. In that paper, no model has been proposed. To our knowledge, the model proposed in our paper is the first model of this particular phenomenon. For that reason, we are unable to compare our model to any other model.

The model is not intended to provide a new approach to intrusion detection. Instead, it can be used to estimate the likely variations in work load related to a given network, a useful information for a MSSP manager. It also provides a rough indication of how many of the network's intrusions are of the benign "soft" nature as opposed to hard-to-detect, more dangerous "hard" intrusions. In that sense, it could be used towards a metric of risk, and for characterizing the nature of threats to the network.

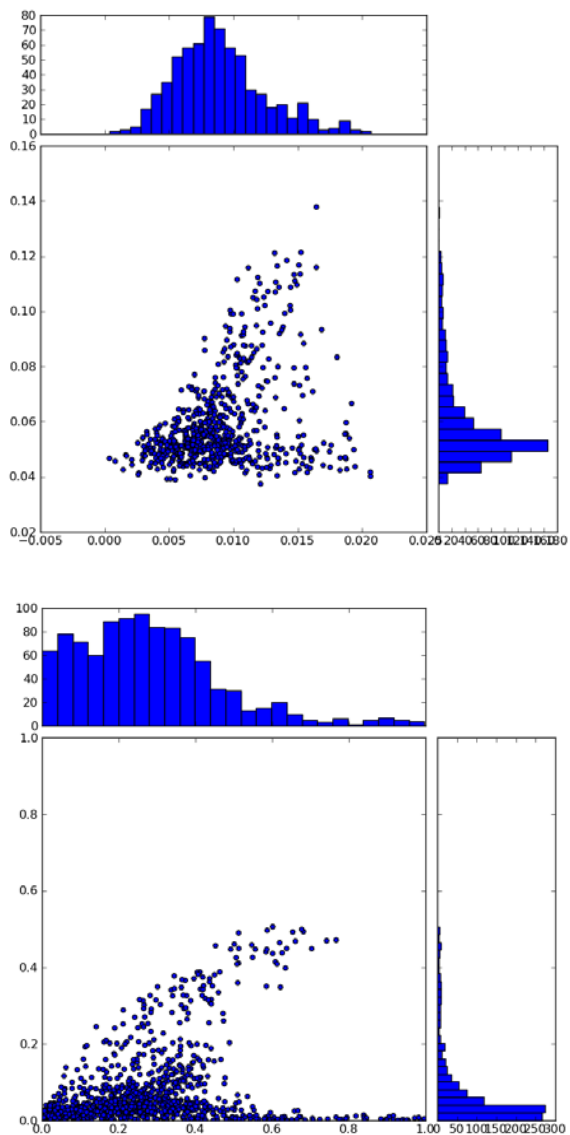

Fig. 4. Joint plots for rate parameters (top) and switching parameters (bottom) for the two-state model of a single network. Rates are given against an arbitrary time scale; probabilities are rendered as fit.

VII. CONCLUSIONS

To the best of our knowledge, this is the first demonstration, with strong empirical evidence that involves data from multiple organizations and use of several statistical techniques, of burstiness in the process of detecting infections effected by cyber threats against networks of large organizations. Testing the arrival and inter-arrival times of the time series via the Kolmogorov test shows that in three of the five organizations, this distribution clearly violates a Poisson process, with p-values of less than 0.001, while a fourth organization has weaker p-values of <0.05 for each. The final organization has a p-value of 0.02 for the distribution of arrival times, however the distribution of interarrival times is not significant at a p-value of 0.222. Examining a variant of the K-statistic across a range of scales shows that the rate at which observations cluster in all five data sets is significantly higher than that of the null model (Poisson process). Finally, estimating p-values via Monte Carlo simulation for the memory and burstiness parameters shows that three of the five time series display marked deviations from the null model for both statistics. Although the burstiness is less pronounced in some cases, it is highly visible in all tests performed on the two largest datasets that cover the largest number of infections.

However, we do not conclude that burstiness is always present, in a general case, or can be observed in all networks and in all intrusion detection processes. In other words, we only show that burstiness is present in some cases, but not necessarily in all generic cases. It is entirely possible that there may exist classes of networks, and/or associated intrusion processes, and intrusion detection processes where burstiness is either absent in principle, or cannot be detected from the available data. Identifying conditions under which burstiness exists, and can be observed, is a topic of future research.

Burstiness and memory parameters suggest that the nature of the process under considerations is more reminiscent of natural processes such as earthquakes than of anthropogenic processes such as sending emails. Unlike anthropogenic processes, intrusion detection exhibits strong memory. We propose the hypothetical mechanism – the Analyst Knowledge Threshold – of the burstiness we observe in intrusion detection.

We have developed a two-state model that yields plausible values of parameters with which it produces report-generation trajectories closely matching the actually observed data. The model can be used to estimate the likely variations in work load related to a given network, a useful information for a MSSP manager. It also provides a rough indication of how many of the network's intrusions are of the "soft" nature as opposed to "hard", more dangerous malware. In that sense, it could be used towards a metric of risk to the network.

While the simple two-state model presented here provides some insight into the process of network intrusions, it remains a highly idealized and very rough approximation. In particular, it does not account for variation in the rates of undetected spread of internal infections, or relative ease of detection of different variants. In the future, we plan to augment the model with latent variables that can model such infection-to-infection variability with higher fidelity.